\documentclass[sigconf]{acmart}

\usepackage[english]{babel}
\usepackage{color}
\usepackage{xcolor}
\usepackage{framed}
\usepackage{multirow}
\usepackage{rotating}
\usepackage{fp}
\usepackage{xspace}
\usepackage{amsfonts}\usepackage{amsthm}
\usepackage[T1]{fontenc}
\usepackage{listings}
\usepackage{todonotes}
\usepackage{glossaries}
\glsdisablehyper
\usepackage{xfrac}
\usepackage{cleveref}
\crefformat{footnote}{#2\footnotemark[#1]#3}
\usepackage[shortlabels]{enumitem}
\definecolor{lfdblack}{HTML}{000000}
\definecolor{lfdyellow}{HTML}{E69F00}
\definecolor{lfddgrey}{HTML}{999999}
\definecolor{lfdgreen}{HTML}{009371}
\definecolor{lfdhgrey}{HTML}{beaed4}
\definecolor{lfdred}{HTML}{ed665a}
\definecolor{lfdblue}{HTML}{1f78b4}
\definecolor{bggray}{gray}{0.9}

\definecolor{shadecolor}{gray}{0.95}
\definecolor{dimgray}{rgb}{0.3, 0.3, 0.3}
\newboolean{anonymous}
\setboolean{anonymous}{false}
\newcommand{\TODO}[1]{\textcolor{red}{TODO: #1}}

\ifbool{anonymous}{
\renewcommand{\TODO}[1]{}
}{}

    \usepackage[commandnameprefix=always, defaultcolor=magenta]{changes}

\definecolor{goldenbrown}{rgb}{0.6, 0.4, 0.08}

\addto\extrasenglish{%
}

\newcommand{\reprolink}{https://github.com/ReproPackage/Q25}
\newcommand{\ie}{\emph{i.e.}\xspace}
\newcommand{\eg}{\emph{e.g.}\xspace}

\makeatletter
\def\calcLength(#1,#2)#3{%
  \pgfpointdiff{\pgfpointanchor{#1}{center}}%
  {\pgfpointanchor{#2}{center}}%
  \pgf@xa=\pgf@x%
  \pgf@ya=\pgf@y%
  \FPeval\@temp@a{\pgfmath@tonumber{\pgf@xa}}%
  \FPeval\@temp@b{\pgfmath@tonumber{\pgf@ya}}%
  \FPeval\@temp@sum{(\@temp@a*\@temp@a+\@temp@b*\@temp@b)}%
  \FProot{\FPMathLen}{\@temp@sum}{2}%
  \FPround\FPMathLen\FPMathLen5\relax
  \global\expandafter\edef\csname #3\endcsname{\FPMathLen}
}
\makeatother

\newcommand{\rqbox}[2]{
\par\smallskip
  \noindent
  \colorbox{dimgray}{%
  \parbox{\dimexpr\linewidth-2\fboxsep\relax}{%
    \color{white}\bfseries #1%
    }%
    }
  \noindent
  \colorbox{shadecolor}{%
  \parbox{\dimexpr\linewidth-2\fboxsep\relax}{%
    \color{black} #2%
    }%
    }
\par\smallskip
}

\ExplSyntaxOn
\NewDocumentCommand{\providewarningcommand}{m+m}
 {
  \cs_if_exist:NTF #1
   {
    \PackageWarning{custom}{Command~#1~already~exists;~redefining~it}
    \RenewDocumentCommand #1 {m} {#2}
   }
   {
    \NewDocumentCommand #1 {m} {#2}
   }
 }
\ExplSyntaxOff

\providewarningcommand{\enquote}{``#1''}
\providewarningcommand{\censor}{``#1''}
\providewarningcommand{\blackout}{``#1''}

\newcommand{\anonorcid}[1]{
\ifbool{anonymous}
{\orcid{0000-0000-0000-0000}}
{\orcid{#1}}
}
\ifbool{anonymous}{}{\renewcommand{\censor}[1]{#1}}
\ifbool{anonymous}{}{\renewcommand{\blackout}[1]{#1}}

\newcommand{\anonemail}[1]{
\ifbool{anonymous}
{\email{xxx.xxx@xxx.xx}}
{\email{#1}}
}

\ifbool{anonymous}{}{}

\newcommand{\anontodo}[1]{
\ifbool{anonymous}
{}
{\todo{#1}}
}

\newacronym{mde}{MDE}{Model-Driven Engineering}
\newacronym{dsl}{DSL}{Domain-Specific Language}
\newacronym{uml}{UML}{Unified Modeling Language}
\newacronym{mqsf}{MQSF}{Munich Quantum Software Forum}
\newacronym{qce}{QCE}{IEEE International Conference on Quantum Computing and Engineering}
\newacronym{qpu}{QPU}{Quantum Processing Unit}
\newacronym{hpc}{HPC}{High-Performance Computing}
\newacronym{sdk}{SDK}{Software Development Kit}
\newacronym{api}{API}{Application Programming Interface}
\newacronym{qpl}{QPL}{Quantum Programming Language}

\AtBeginDocument{%
  }

\copyrightyear{2026}
\acmYear{2026}
\setcopyright{cc}
\setcctype{by}
\acmConference[MODELS Companion 2026]{ACM/IEEE 29th International Conference on Model Driven Engineering Languages and Systems}{October 04--09, 2026}{Málaga, Spain}
\acmBooktitle{ACM/IEEE 29th International Conference on Model Driven Engineering Languages and Systems (MODELS Companion 2026), October 04--09, 2026, Málaga, Spain}
\acmDOI{10.1145/3837062.3839073}
\acmISBN{979-8-4007-2903-4/2026/10}

\begin{document}


\title[Know Your Qubits, Know Your Users: Personas for Quantum Software]{Know Your Qubits, Know Your Users:\\ Personas for Quantum Software}


\settopmatter{authorsperrow=4}

\blackout{\author{Lukas Schmidbauer}}
\authornote{Both authors contributed equally to this research.}
\anonemail{lukas.schmidbauer@othr.de}
\anonorcid{0009-0001-7171-0865}
\affiliation{%
  \blackout{\institution{Technical University of \hbox{Applied Science Regensburg}}}
  \blackout{\country{Germany}}
}

\blackout{\author{Joshua Ammermann}}
\authornotemark[1]
\anonemail{joshua.ammermann@kit.edu}
\anonorcid{0000-0001-5533-7274}
\affiliation{%
  \blackout{\institution{Karlsruhe Institute of Technology}}
  \blackout{\country{Germany}}
}

\blackout{\author{Laura Schulz}}
\anonemail{lschulz@anl.gov}
\anonorcid{0000-0002-4702-3440}
\affiliation{%
  \blackout{\institution{Argonne National Laboratory}}
  \blackout{\country{USA}}
}
   
\blackout{\author{Jose Garcia-Alonso}}
\anonemail{jgaralo@unex.es}
\anonorcid{0000-0002-6819-0299}
\affiliation{%
  \blackout{\institution{University of\\Extremadura}}
  \blackout{\country{Spain}}
}

\blackout{\author{Robert Wille}}
\anonemail{robert.wille@tum.de}
\anonorcid{0000-0002-4993-7860}
\affiliation{%
  \blackout{\institution{Technical University of Munich}}
  \blackout{\country{Germany}}
}

\blackout{\author{Sebastian Feld}}
 \anonemail{S.Feld@tudelft.nl}
 \anonorcid{0000-0003-2782-1469}
 \affiliation{%
   \blackout{\institution{Delft University of Technology}}
   \blackout{\country{Netherlands}}
}

\blackout{\author{Ina Schaefer}}
\anonemail{ina.schaefer@kit.edu}
\anonorcid{0000-0002-7153-761X}
\affiliation{%
  \blackout{\institution{Karlsruhe Institute of Technology}}
  \blackout{\country{Germany}}
}

\blackout{\author{Wolfgang Mauerer}}
\anonemail{wolfgang.mauerer@othr.de}
\anonorcid{0000-0002-9765-8313}
\affiliation{%
  \blackout{\institution{Technical University of \hbox{Applied Science Regensburg}}}
  \blackout{\country{Germany}}
}
\additionalaffiliation{%
  \blackout{\institution{Siemens AG, Foundational Technologies}}
  \blackout{\country{Germany}}
}

\renewcommand{\shortauthors}{\censor{Schmidbauer et al.}}

\begin{abstract}
%
The advancement of quantum hardware and the intricacies of quantum computing make well-designed quantum software increasingly necessary.
Due to the interdisciplinarity of the field, it is crucial to understand the perspectives and specific needs of involved stakeholders, for example, to balance the desired level of abstraction with the exposition of (hardware)-specific details.
%
In this work, we conduct a stakeholder-based analysis to identify personas of quantum software as a means of creating meaningful, user-tailored quantum software.
We conducted an expert focus group at a Dagstuhl seminar in 2024 and qualitative interviews with practitioners at conference IEEE QCE in 2025, from which we derive eleven personas of potential users and stakeholders for quantum software. 
We discuss these personas regarding their use cases, interests, constraints and abstraction level.
%
%
\end{abstract}



\received{20 February 2007}
\received[revised]{12 March 2009}
\received[accepted]{5 June 2009}

\maketitle

\section{Introduction}
\label{sec:intro}

Quantum computing, as a continuously developing area of research, promises computational or other advantage over classical approaches~\cite{shorPolynomialTimeAlgorithmsPrime1997, Aaronson_2026, Eisert2025}.
Researchers are lowering the bar for achieving practical advantages in quantum agnostic problems with recent advances in mappings to decoding problems~\cite{Jordan_2025, Khattar2025, Gu2025, Chailloux2025, Sabater2025}.

Exploiting the advantages of
maturing quantum hardware~\cite{deLeon2021} requires the development of practical quantum software, which has become a research focus of its own ~\cite{zhaoQuantumSoftwareEngineering2021, carbonelliChallengesQuantumSoftware2024a, murilloQuantumSoftwareEngineering2025}.
One of the key issues in quantum software engineering is designing languages and, in particular, abstractions that can make quantum computing accessible to users from many domains without direct relation to quantum.
While languages and frameworks have become available ~\cite{ferreiraExploratoryStudyUsage2025a},
it is common practice to incorporate quantum-specific constructs into existing general-purpose programming languages (\eg, Python).
For example, the widely adopted frameworks Qiskit and Cirq employ this approach~\cite{upadhyayAnalyzingEvolutionMaintenance2025},
albeit special-purpose languages have also been proposed~\cite{bichsel2020silq,svore2018qsharp,Mauerer:2005}.
Both offer an abstracted view on hardware and easy deployment of algorithms expressed as quantum circuits, which benefits higher level applications and users with programming experience. 
Conversely, for some other users, hardware-specific details (\eg, error-rates, topology, qubit type) leverage the (application dependent) optimization potential of available systems (\eg, Co-Design approaches can use exposed details \cite{Hila2023,Li2021,Safi2025}).
Thus, balancing the level of abstraction and exposition of details is a non-trivial, but necessary task in the development of quantum software.

\Gls{mde} applied to quantum software~\cite{gemeinhardtModelDrivenQuantumSoftware2021a, aliModelingQuantumPrograms2020} supports the software development process by leveraging models as central artifacts.
One major advantage of \Gls{mde} is that models are created for a specific purpose and stakeholder~\cite{vaPlterModelDrivenSoftwareDevelopment2013}.
Thus, \Gls{mde} can provide domain experts with tailored \Glspl{dsl} that expose relevant details of quantum software in a familiar way.
Researchers in the field of quantum mechanics, for example, could benefit from an interface that allows them to express Hamiltonians in line with their mental models.
The work of \citeauthor{murilloQuantumSoftwareEngineering2025}~\cite{murilloQuantumSoftwareEngineering2025} gives a comprehensive overview of current quantum \Gls{mde} research.
Ongoing work includes defining domain-specific meta-models for quantum circuits~\cite{aliModelingQuantumPrograms2020, gemeinhardtModelDrivenFrameworkCompositionBased2023} and the high-level Qrisp quantum programming language~\cite{bockDesigningMetaModelEclipse2025a}, and extending the widely used \Gls{uml} for quantum computing~\cite{perez-castilloDesignClassicalquantumSystems2022}.

As an emerging technology spanning a wide range of disciplines, quantum computing involves a variety of domain experts, who could benefit from representations tailored to their needs, such as \Glspl{dsl}.
Involved stakeholders range from end users who want to solve problems with quantum computing to engineers who are building the necessary quantum hardware.
This interdisciplinarity becomes particularly evident in the field of quantum software, as exemplified by major quantum software events, such as the \Gls{mqsf}~\cite{Wille_2025_MQSF}, where the number of attendees with a background in physics still exceeds those with a background in computer science.
Therefore, to develop quantum software tailored to stakeholders, it is essential to understand their specific needs and perspectives.
Although those working on different parts of quantum software each consider a specific portion of stakeholders (relevant for their project), knowledge about these stakeholders is fragmented across the various workplaces, institutions and smaller specialist communities involved \cite{feld2025full}.

To address the lack of consolidated stakeholder characterizations, we conduct a \textit{persona-based analysis} to identify and characterise stakeholder groups for quantum software.
We use \textit{personas}~\cite{cooperFaceEssentialsInteraction2014}---a method from interaction design---to define user groups.
Personas support practitioners and researchers across domains in the design process and have been widely adopted and adapted~\cite{purittPersonasPracticeAndTheory2003, changPersonasTheoryPractices2008, salminenUseCasesPersonas2022}.
They provide a model for thinking and communicating about user groups, serving as a common language and facilitating communication.
The quantum \Gls{mde} community can benefit from personas, as they provide information on who the tailored models and \Glspl{dsl} are being crafted for.
In general, personas are used to design meaningful interfaces that users find satisfactory.
Future practitioners may adopt the Goal-Directed Design process~\cite{cooperFaceEssentialsInteraction2014} (see \autoref{sec:Background}) in its entirety when designing quantum software products.
Therefore, readily available personas for quantum software support the ongoing initiative to create purposeful, engaging quantum software that meets the needs of many different stakeholders.
More precisely, to advance towards this goal, in this work we:
\begin{itemize}
    \item Present results from two studies we conducted at different venues to derive personas for quantum software: 
    \begin{enumerate}[(a)]
        \item To incorporate the expertise of the specialists in the field, we conducted a focus group with quantum software engineering experts at the Dagstuhl Seminar 24512 in 2024 (see \autoref{sec:personas}). 
        The focus group yielded five short-, six mid-, and five long-term personas.
        \item To incorporate the perspective of practitioners actively working on quantum software, we conducted qualitative interviews with practitioners at IEEE \acrshort{qce} 2025 (see \autoref{sec:Results}).
        The interviewees identified nine personas, focusing on the short-term perspective.
    \end{enumerate}
    \item Aggregate 11 personas for quantum software from the insights gathered in the studies (see \autoref{sec:discussion}).
\end{itemize}

\section{Background: Personas in Interaction Design}
\label{sec:Background}

\label{subsec:GDD}
\textit{Interaction Design} is concerned with identifying a product's core users and aims to create efficient and appealing design solutions by understanding users and their goals.
In this context, Cooper introduced 
the \textit{Goal-Directed Design}~\cite{cooperFaceEssentialsInteraction2014} process, which uses \textit{personas} for thinking and communicating about user groups.

In Goal-Directed Design, three types of models are distinguished: user models (\ie{}, conceptional or mental models), represented models, and implementation models. 
Implementation models reflect technology, whereas user models reflect a user's vision, hiding the complex mechanisms they do not need to know or care about. 
The represented models lie between the designers and the users, capturing how the designers display an application's functionality. 
Represented models that are closer to the user models are easier for them to understand.

\begin{figure}[bt]
    \centering
    \includegraphics[width=0.92\linewidth]{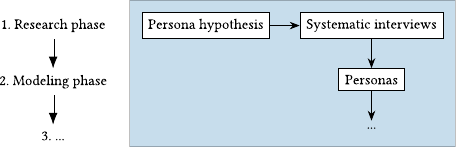}

    \caption{First phases (left) and corresponding activities (right) of the Goal-Directed Design process adapted from~\cite{cooperFaceEssentialsInteraction2014}.
    Personas are used in later phases to design user interfaces.
    }
    \Description{Text book background: Persona Hypothesis then systematic interviews and then arriving at personas.}
    \label{fig:gdd}
\end{figure}

The Goal-Directed Design process has six phases, of which we show the first two in \autoref{fig:gdd} (left).
The initial \textit{research phase} aims to collect qualitative data on potential or actual users of the product.
To achieve this, systematic field study techniques are employed, such as user observations and interviews.
Interviews are typically conducted one-on-one with broad, open-ended questions, lasting no more than an hour in early interviews, as they focus on gathering domain knowledge from the user’s perspective.
In this phase, \textit{persona-hypotheses} (see \autoref{fig:gdd}: right) are created based on information from subject matter experts, stakeholders and literature reviews, and serve
as an initial attempt at defining the different kinds of users.
The outcome of this phase is a set of behaviour patterns that help categorise modes of use of a potential or existing product.
In technical domains, they often map to professional roles and are the basis for personas in the modelling phase.

The \textit{modeling phase} uses \textit{personas} as a user model (see \autoref{fig:gdd}), synthesised from previously acquired patterns.
Personas (\ie, descriptive models) provide a precise way of thinking and communicating about users.
They are based on behaviours and motivations from real persons, and capture users' behaviour, thinking, goals and reasoning.
While personas represent groups of users, they are presented as individual humans and thus have a distinct narrative structure.
Hence, they provide a unique common language between stakeholders, developers, and designers.
In contrast, other user models
do not attempt to convey broader human motives and contexts.
Personas focus on a specific domain and product, so only personas based on research covering similar contexts should be reused.

In the following phases of the Goal-Directed Design process, the defined personas are used to design meaningful interfaces which are satisfactory for the users.
We see immediate benefits in using personas for quantum software, in that they serve as a common language.
Though, future practitioners may fully adapt the Goal-Directed Design process to design quantum software products.

\section{Research Questions and Methodology}
\label{sec:ResearchAndMethodology}

When searching for meaningful quantum software models and \Glspl{dsl}, it is important to consider that they are designed for a particular purpose and user.
In the interdisciplinary field of quantum software, it is particularly important to understand the perspectives and specific needs of the stakeholders involved.
To better understand these, this work addresses the following research questions:
\begin{description}
    \item[\textbf{RQ1}] Who are the relevant stakeholders in quantum software,
both now and in the future (short-term, mid-term, and long-term)?
    \item[\textbf{RQ2}] What are the specific needs of stakeholders in quantum software, and how can abstraction help to reach their goals?
\end{description}

Stakeholders are interested in particular use cases, which are driven by specific interests (\ie, the key objectives they wish to achieve).
Constraints or side conditions apply to how objectives can be reached. 
Different levels of abstraction are required depending on the use case and will vary over time as the capabilities of machines improve.
Hence, we derived sub-research questions of 
\textbf{RQ2}:
\begin{description}
    \item[\textbf{RQ2.1}] What use cases for quantum software do these personas have? Which problems do they want to solve?
    \item[\textbf{RQ2.2}] What interests do these personas have in using quantum software?
    \item[\textbf{RQ2.3}] Which constraints (side conditions) for quantum software and their use cases do these personas have?
    \item[\textbf{RQ2.4}] What are the abstraction levels (\eg, language building blocks and idioms) that these personas want to use?
\end{description}

\textbf{RQ2.1} regards use cases as (functional) stakeholder goals, which the stakeholders want to achieve using quantum software.
In \textbf{RQ2.2}, broader motives driving these stakeholders, such as business motives and the pursuit of scientific insight, are investigated.
\textbf{RQ2.3} considers constraints, meaning limitations or challenges, that stakeholders face when engaging with quantum software.
\textbf{RQ2.4} aims to identify which abstractions the stakeholders want to interact with quantum software with respect to their mental models. Abstractions in a quantum software stack can range from domain specific descriptions (high-level), over implementation and compilation using programming languages and frameworks, to the low-level physical realisation and integration of (quantum) hardware.

Goal-Directed Design (see \autoref{subsec:GDD}) is a widely adopted way to investigate user groups when designing software.
It uses the concept of \textit{personas} as descriptive models of users, presented as individual humans.
Personas are an established methodology~\cite{purittPersonasPracticeAndTheory2003, changPersonasTheoryPractices2008, salminenUseCasesPersonas2022} that we propose to use when designing quantum software products.
For this work, we do not have a concrete product in mind, but want to give a general overview of possible kinds of users for quantum software.
Thus, we do not define concrete personas, but formulate a \textit{persona-hypothesis} as an initial step at defining the different kinds of users according to the Goal-Directed Design process.
Future practitioners creating quantum software products may use our persona-hypothesis as a basis for initial interview planning, to then create concrete personas for their product.

To derive meaningful personas, we apply qualitative methods suitable for an exploratory setting, allowing participants to respond more freely~\cite{lenbergQualitativeSoftwareEngineering2024}.
To build personas on the experiences from experts as well as practitioners, we conduct two separate studies.
First, we conduct a focus group with the experts of the \emph{Software Abstraction in Quantum Computing} working group~\cite{ali_et_al:DagRep.14.12.63}, as part of a Dagstuhl seminar in 2024 (see \autoref{sec:personas}).
This seminar provides a rare opportunity to interact face-to-face with a range of experts in quantum software engineering.
Second, to also incorporate the perspective of practitioners actively working on quantum software, 
we conduct a qualitative interview study with relevant stakeholders
(see \autoref{sec:Results}) at IEEE \acrshort{qce} 2025, an event attended by academics and industry professionals~\cite{QCE_2026_Attend}. 
To answer \textbf{RQ1} and \textbf{RQ2}, we then aggregate the insights from both studies as a persona-hypothesis\footnote{
For the sake of brevity, we may refer to the different kinds of users according to our proposed \textit{persona-hypothesis}
as \textit{personas} in the remainder of this work.} in \autoref{sec:discussion}.
\section{Personas by Expert Focus Group at Dagstuhl}
\label{sec:personas}
We conducted a focus group with the experts from the \emph{Software Abstraction in Quantum Computing} working group ~\cite{ali_et_al:DagRep.14.12.63}, in which authors of this work participated, as part of the \href{https://www.dagstuhl.de/24512}{Dagstuhl Seminar 24512} (link in pdf).
In December 2024, the seminar brought together 24 experts in quantum software engineering for five days. 
Of these, three days were dedicated to discipline-specific working group discussions.
The following results\footnote{Take into consideration that we paraphrase. We explicitly do not characterise people.} were produced by an expert group of eight to ten people, with attendees varying across individual sessions.

\begin{table*}
    \centering
    \caption{Personas of quantum software according to the Dagstuhl expert group, categorised by quantum computing hardware availability in the short-, mid- and long-term.}
    \label{tab:user_groups_time}
    \begin{tabular}{p{0.27\textwidth}p{0.27\textwidth}p{0.37\textwidth}}\toprule
            \textbf{Short-term} & \textbf{Mid-term}& \textbf{Long-term} \\ \midrule
            Quantum curious (business) & Business users & Business application engineer (domain expert)  \\
            \acrshort{hpc} engineer (\enquote{Offloading}) & \acrshort{hpc} engineer (\enquote{QC at scale}) & \acrshort{hpc} engineer (\enquote{QC at large scale}) \\
            Embedded quantum developer & Embedded quantum developer & Embedded quantum developer (\enquote{power user}) \\
            Quantum algorithm researcher & Quantum algorithm developers (CS) & Quantum algorithm developers (CS) \\
            Quantum mechanics researcher & Quantum mechanics researcher & Quantum mechanics researcher / physicist \\
           & Material scientist & Material scientist\\
           & Early adopters & \\\bottomrule
    \end{tabular}
\end{table*}

According to the expert group, improving quantum hardware is expected to lead to a diversification of the user base over time.
Quantum computers featuring only tens to hundreds of physical quantum bits are already available in the \textit{short-term}. 
This means that execution is fundamentally limited to log-depth circuits~\cite{Preskill:2018, Greiwe2023}. 
Furthermore, the interconnects necessary to create distributed \Glspl{qpu} are lacking.
In the \textit{mid-term}, experts project systems with thousands to hundred of thousands of physical qubits and error-correcting codes. 
These codes may not fully correct all potential errors, but still enable accurate execution of circuits beyond log-depth with a reasonable success rate.
Rerunning a calculation may be necessary if too many errors occur, but this limitation is not decisive for practical utility. 
Initial interconnects to combine tens to hundreds of \Glspl{qpu} are available through hardware vendors aiming to create first of such systems within this decade, according to their roadmaps \cite{Swayne_2025}.
A \textit{long-term} goal for quantum hardware is to de-risk physics in the sense that systems can be scaled using multi-\Gls{qpu} setups with reliable physical interconnects that transmit quantum information faithfully across \Glspl{qpu}.

To address \textbf{RQ1}, the experts created personas at the \textit{short-}, \textit{mid-} and \textit{long-term} time intervals representing expected stages of technological maturity.
The personas involved in quantum software within the time intervals that were identified in the expert group are summarised in \autoref{tab:user_groups_time}.
They identified five short-, seven mid-, and six long-term personas\footnote{Note that we expect profiles of stakeholders to change as hardware matures.} covering a variety of domains:
\paragraph{Short-term}
Initially \textbf{curious} (business) \textbf{users} can conduct pilot studies serving as a starting point for scalable quantum operations.
Knowledge of concrete pipeline stages (\ie, transformations) and quantum specific details is required~\cite{Thelen2024, Schmidbauer2023}.
Some quantum learners in the short-term are interested about quantum computing by conducting exploratory studies. 
Hence, they favour graphical, intuitive, interactive and explaining interfaces not only in circuit level abstraction, but also in IDE's and when applying quantum software idioms.
Business users operate on a high abstraction level. \textbf{\Gls{hpc} engineers}~\cite{Elsharkawy2024} and \textbf{embedded quantum developers}~\cite{Ramsauer2025} are close to system integration of hardware to (a) develop scalable and distributed quantum computing and (b) develop the necessary low-level interfaces that make system-level integration of quantum computing possible.
\textbf{Quantum algorithm researchers} also work on interfaces, but at higher levels of abstractions---ranging from purely logical quantum algorithms (not limited to circuit representation) to hardware-aware (\eg, for topology, noise, operations) quantum algorithms. 
To improve the abilities and efficiency of known algorithms, quantum algorithm researchers in the short-term operate on low-level abstractions (\ie, gates and unitaries), as well as on high-level languages describing the functional behaviour of algorithms.
The goal for \textbf{quantum mechanics researchers} is scientific knowledge gain, which in the short-term is primarily based on physics. 
Similar to business users, they are able to conduct (limited) pilot studies for their use cases.
They are more specialised in the simulation of various quantum systems, while being constrained by the need to translate to circuit level.
Accuracy of simulations and discovery of (new) phenomena are major interests.
\paragraph{Mid-term}
The availability of many (mostly) error-corrected qubits enables \textbf{business users} to scale their pilot studies to industry relevant problem sizes. 
While the implementation still necessitates an understanding of quantum-specific intricacies, this requirement is less pronounced compared to pilot studies conducted in the short-term.
Although \textbf{\Gls{hpc} engineers} and \textbf{embedded quantum developers} are still close to system integration, their tasks shift from integrating and optimising single \Glspl{qpu} to multiple (quantum mechanically connected) \Glspl{qpu} \cite{escofet2025revisiting}.
When considering the application of large scale, heterogenous calculations with abstractions in line with, for example, \href{https://www.sandia.gov/ccr/software/lammps/}{LAMMPS} (link in pdf), the ability to interleave communication and coordinate calculations is a major constraint for \Gls{hpc}.
As systems scale, quantum algorithms are subject to different (performance) metrics (\eg, the limitation to log depth circuits is gradually relaxed), which entails a broader view on computer science paradigms and complexity theoretic consideration for \textbf{quantum algorithm developers}.
Also, for \textbf{quantum mechanics researchers}, computer science becomes a necessary field of expertise to efficiently execute experiments on multiple connected \Glspl{qpu}.
A portion of mid- and long-term quantum mechanics researchers, particularly high energy physicists, are interested in Schwinger models or lattice gauge simulations~\cite{Franz2024} with attention to accuracy and new phenomena. 
In line with Lagrangians and Hamiltonians as their main abstractions, they are constrained by the expressibility of low code and maths\footnote{For instance, Hamiltonian simplifications and their expressibility in frameworks.}.
Due to the increased availability of qubits, simulations of materials (\eg, multi-atom simulations, macroscopic property simulations) become feasible, which opens the field of quantum computing for \textbf{material scientists} to develop new and optimise existing materials and compounds~\cite{McArdle2020,Franz2026, Wulff2024}.
Concurrently, \textbf{early adopters} use quantum computing in production settings.
\paragraph{Long-term}
As scalable hardware becomes available, \textbf{business users} use quantum computers on industrial scale and become abstract from pipeline and quantum specific details, while focusing on domain expertise.
While a subset of long-term \textbf{business application engineers} are driven by financial benefits, they are interested in result quality, runtime and customer value in a quantum agnostic abstraction, meaning that the execution on a quantum computer is abstracted away from the user, which enables similar (classical) interfaces as before.
\textbf{\Gls{hpc} engineers} integrate multiple (distant) quantum computers at large scale.
Although there exists a rich environment of (optimising) tools and toolchains in the long-term, \textbf{embedded quantum developers} can leverage their system specific expertise to realise systems full potential.
In the short- and mid-term, \textbf{quantum algorithm developers} are focused on satisfying hardware constraints and optimise resource requirements to be able to execute algorithms. 
As systems scale in the long-term, additional metrics (\eg, ease of use, customisability, (response) time, cost-effectiveness, political restrictions, security, data privacy, protection of intellectual property) become more pronounced around which new algorithms are designed.
Similar to mid-term considerations, \textbf{quantum mechanics researchers} are able to execute large scale problems, but need to consider scaling effects, approximation quality and solution guarantees.
Also, \textbf{material scientists} achieve scalability comparable to quantum systems, enabling new methods (\eg, material interaction simulation on macroscopic scale), which may also make biological simulations possible~\cite{Needleman_2017}.

\section{Personas by Practitioner Interviews at QCE}
\label{sec:Results}

To gather insights from practitioners, we conducted an interview study with relevant stakeholders.
Interviews were conducted as one-on-one interviews with broad open-ended questions, focusing on gathering domain knowledge from the interviewees’ perspective.
We conducted our study as a workshop session at the \Gls{qce} in 2025.
Before the interviews, the interviewees received a \href{https://doi.org/10.5281/zenodo.21837115}{brief presentation} (link in pdf) about workshop goals and involved personas, and completed a demographic questionnaire.

\paragraph{Demographic Questionnaire}

\begin{figure}[tb]
\centering
\includegraphics[width=0.92\linewidth]{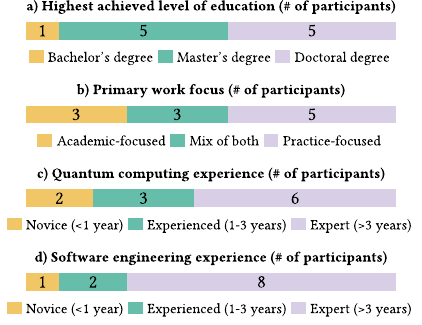}
        
\caption{Demographics of \acrshort{qce} 2025 interviewees ($N=11$).}
\Description{Stacked bar charts, each displaying the responses of participants to a question from the demographic questionnaire.}
\label{fig:demographics_success}
\end{figure}

We asked our participants a series of demographic questions to confirm that they were experienced practitioners from academia and industry.
We asked four demographic questions regarding the participants level of education, work focus, and experience in quantum computing and in software engineering. 
The questions were provided through \href{https://docs.google.com/forms/d/e/1FAIpQLSeK4Q85gtfXlPvbtb_KswJ1Is9FjXfJGZQ9UDXpNA4gW5etbw/viewform?usp=publish-editor}{Google Forms} (link in PDF).
We obtained demographic information for eleven \Gls{qce} participants, shown in \autoref{fig:demographics_success}.
Higher education was common: Five participants held a Master's, and another five a doctoral degree (\autoref{fig:demographics_success} a).
The primary work focus spanned from academic over mixed to practical-focused, leaning towards  practical-focused (\autoref{fig:demographics_success} b).
Quantum computing and software engineering experience of participants (\autoref{fig:demographics_success} c, d) are similar, with mostly expert participants.

\newcommand{\layercolorof}[1]{%
  \ifcase#1 lfdyellow\or lfdgreen\or lfdhgrey\or lfdred\or lfdblue\fi}

\paragraph{Practitioner Interviews}
To allow for more detailed discussions,
we conducted single person interviews (one interviewee, one interviewer, and one note-taker) in multiple sessions of our workshop \emph{Care and Maintenance of Quantum Software Stakeholders} at IEEE \Gls{qce} 2025. 
In total, there were three interviewers and three note-takers (all with academic background). 
The interviews were semi-structured, using guiding questions to allow for flexible conversation and gather domain knowledge from the interviewee’s perspective (see \autoref{subsec:GDD}).
We asked interviewees to share their insights on personas, encouraging them not only to consider those defined in \autoref{sec:personas}, but also to introduce additional ones when appropriate.
To guide the conversation, we use \textbf{RQ1}, \textbf{RQ2} and \textbf{RQ2.1-RQ2.4} as interview guiding questions.
Interviews were voluntary --- also after completing the demographic questionnaire, and hence we conducted 9 interviews
, lasting between 15 and 40 minutes. 
Due to privacy considerations, we cannot offer raw data of each interview.
\paragraph{Interview Results}
The interviewees identified nine personas (\textbf{RQ1}).
In comparison to \autoref{tab:user_groups_time} and \autoref{sec:personas}, business users and researchers were described in depth by the participants and categorized into short-, mid- and long-term.
Participants described the business users as a combination of short-, mid- and long-term users, with the key differences in the extent of their application.
However, the common factor across all is the profit-driven mindset that motivates decisions and actions.
The described researchers (in varying degrees) cover short-, mid- and long-term by (among others) applying quantum algorithms to existing hardware, making pilot studies on industrial use cases for mid-term applications and considering long-term effects on security.
They cover a broad spectrum of tasks and hence distinctions to other personas are not sharp.
Participants further identified the personas of physicists, chemists, simulation focused engineers, quantum algorithm designer, \Gls{hpc} engineers, platform builders / developers, and government contractors, but did not go into detail about how they will change over time.
In the following, we provide paraphrased insights into personas (\textbf{RQ2}) described by participants:

\textbf{Business users} are relatively high-level personas, described by three participants.
Although they are primarily focused on maximising profit (RQ2.2), they have varying use cases ranging from (a) acquiring new customers, (b) protecting intellectual property, which also considers security aspects of quantum computing, to (c) driving research into industry applications (RQ2.1). 
Also, one participant mentioned an explicitly longer term business user who has to decide how and to what degree quantum computing benefits a certain problem (\ie, selecting the share of quantum in an (established) process) and thus needs (resource) estimations rather than hardware executable representations (\eg, circuits).
In general, the described business users want to abstract from syntax and implementation details (RQ2.4) and use resource estimates to inform decisions, making them dependent on the accuracy and engineering effort for those estimations (RQ2.3). 
They also favour natural language over rigorous formal languages (RQ2.4).

Two participants
described \textbf{researchers}.
Although~--~in contrast to business users~--~their work is academic, (in-)direct connections to industry exist:
Industry relevant papers in quantum computing not only lay foundations for industrial utility, but also open the field for research on industry applications. 
The researcher personas want to show quantum advantage (RQ2.1) and publish findings (RQ2.2), which is usually fast paced and thus requires a fast prototyping environment (RQ2.3). 
Furthermore, time constraints (of operations) and personnel restrictions (\ie, modular, low-granularity tasks suitable for individual execution rather than large, multi-person work units) are important properties to researchers (RQ2.3).

\textbf{Physicists} were described thrice.
Although sometimes associated with lower level details, interviewees describe them as use case oriented. 
Interests include (a) verifying theories, (b) finding \enquote{new stuff}, (c) exploring condensed matter or (d) simulating problems (\eg, in the aerospace domain), which make them similar to the business user (RQ2.1, RQ2.2). 
They show particular awareness of physical limitations (which is true for any persona, but with less importance).
Another important constraint is the manual transfer of physical details onto, for instance, (quantum) algorithms --- highlighting the importance of automation (RQ2.3).
Their levels of abstraction cover programming languages (\eg, Python) over (quantum) instructions to properties of quantum mechanics, resulting in a lower level of abstraction than for business users (RQ2.4).

Analogously, \textbf{chemists} are interested in solving hard chemistry problems in varying disciplines (\eg, physical, (in)organic and biochemistry problems; RQ2.1, RQ2.2). 
They were mentioned by one participant.
Similar to the physicist, they are constrained by the laws of physics and chemistry, as well as resource limitations, which results in poor scalability of simulations (RQ2.3).
\textbf{Simulation focused engineers} were mentioned by the same participant.
They are interested in problems representable as Hamiltonians (RQ2.1, RQ2.2) and are constrained by the capabilities of hardware (RQ2.3).

\textbf{Quantum algorithm designers} (potentially hardware-agnostic) were described by two participants.
They want to execute an algorithm on multiple backends (RQ2.1) and depend on backend-specific details (to be made available by a software stack; RQ2.3).

\textbf{\Gls{hpc} engineers} focus on quantum-classical integration on multiple (classical) (connected) backends (RQ2.1, RQ2.2). 
They are thus constrained by communication time (\eg, real time connections; RQ2.3) and were mentioned by one participant.

Two participants
describe the \textbf{platform builders / developers} that build quantum platforms (\eg, \Glspl{sdk}, \Glspl{api} and
abstraction frameworks) to benchmark quantum hardware for use cases (RQ2.1, RQ2.2). 
They are constrained by modularity and replaceability of (key) components in a toolchain (\eg, transpilers), by scalability of hardware and lack of standardisation in software (RQ2.3). Constraints include (self-)defined interfaces and the balance between sufficient abstraction and expert access to details (RQ2.4).

One participant
describes \textbf{government contractors}.
In distinction to previous personas, they work on strict security requirements and cannot rely on black-box libraries (RQ2.2). 
They need traceable and deterministic guarantees about software behaviour (RQ2.3).
\section{Aggregated Personas for Quantum Software}
\label{sec:discussion}
The preceding sections presented findings primarily from the perspectives of interviewees and subject-matter experts. 
In this section, we aggregate personas from these findings and complement these perspectives by incorporating our own interpretation of the data.
This paper aims to better understand users of quantum software, so interpretations and previously introduced data may be extended. 

Regarding \textbf{RQ1}, the expert group and \Gls{qce} interviewees introduced personas for users of quantum software.
In particular, the expert group (see \autoref{sec:personas}) considered business users, early adopters, quantum mechanics researchers, material scientists, quantum algorithm researchers, \Gls{hpc} engineers and embedded quantum developers. 
Our interview study (see \autoref{sec:Results}) gives a different perspective: 
Although interviewees described personas with a significant overlap (but different names), they also considered different users (\ie, government contractor, simulation focused engineer, researcher and platform builder). 
\begin{figure}[tb]
    \centering
    \includegraphics[width=0.95\linewidth]{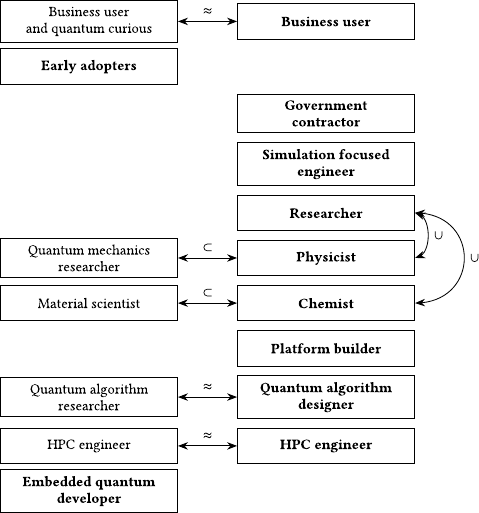}

    \caption{Mapping of Dagstuhl personas (left; \autoref{sec:personas}) to interview personas (right; \autoref{sec:Results}). $A \subset B$ means $A$ was described more specialized than $B$. $A \approx B$ means $A$ and $B$ were described in approximately same scope. 
    Aggregated personas in bold.}
    \Description{Personas as nodes connecting related personas with edges.}
    \label{fig:DagMapsInterview}
\end{figure}
\autoref{fig:DagMapsInterview} compares personas from the Dagstuhl expert group (left) to personas from interviews (right) by showing approximate equivalence ($\approx$) and broader descriptions (\ie, subsets $\subset$).
In particular, we found that descriptions of business users, quantum algorithm designers, and \Gls{hpc} engineers are roughly equivalent (see $\approx$-edges).
Therefore, we considered them as one persona. 
Additionally, interviewees described physicists and chemists more generically than the Dagstuhl personas of quantum mechanics researchers and material scientists (see $\subset$-edges).
We propose to use the physicist and chemist personas in this broad persona-hypothesis context. 
Nevertheless, (quantum) mechanics researchers and material scientists could be considered when deriving concrete personas.
Furthermore, interviewees mentioned supplementary personas, such as government contractor, simulation focused engineer, and platform builder, which originate from practically relevant fields.
In particular, the researcher persona was mentioned as a broad persona (\ie, physicist and chemist can be considered subgroups; see $\subset$-edges). 
Early adopters and embedded quantum developers were only mentioned by the expert group.

\begin{table*}[p]
    \centering
    \caption{Aggregated personas for quantum software as a generalisation from concrete stakeholders, grouped by application and hardware orientation: application oriented personas (business user, early adopter, government contractor), application and hardware oriented personas (simulation focused engineer, researchers, such as  physicist and chemist) and hardware oriented personas (platform builder, quantum algorithm designer, \acrshort{hpc} engineer, embedded quantum developer).
    For each persona we aggregated their use case, interests, constraints and abstraction level according to RQ2.1-RQ2.4 (see \autoref{sec:ResearchAndMethodology}).
    }
    \label{tab:user_groups_zoom}
    {\raggedright\begin{tabular}{p{0.02\textwidth}p{0.09\textwidth}p{0.16\textwidth}p{0.18\textwidth}p{0.2\textwidth}p{0.2\textwidth}}\toprule
            & \textbf{Persona} & \textbf{Use case (RQ2.1)} & \textbf{Interests (RQ2.2)} & \textbf{Constraints (RQ2.3)} & \textbf{Abstraction level (RQ2.4)} \\ \midrule
            \multirow{3}{*}{
            \rotatebox[origin=c]{90}{\makebox[-20pt][r]{\textit{Application oriented}}}
            } 
            & \emph{Business user} & Scalable (profit-driven) quantum operations & Maximising profit, estimating impact on security & Quantum specific intricacies, resource estimation accuracy & High: focus on domain expertise\\\addlinespace\cline{2-6}\addlinespace
            & \emph{Early adopter} & Adopting new quantum technologies in industry applications & Driving business models and revenue growth & Industry relevance, scalability and adaptability & High: focus on market demands \\\addlinespace\cline{2-6}\addlinespace
            & \emph{Government contractor} & Government tasks with strict (security) requirements & Data confidentiality, integrity, and availability & Traceable guarantees about software behaviour, no black-box implementations
            & Mid to high: focus on traceability and observability of software and of algorithms\\\addlinespace\hline\addlinespace
            \multirow{4}{*}{
            \rotatebox[origin=c]{90}{\makebox[-10pt][r]{\textit{Application and hardware oriented}}}
            } 
            & \emph{Simulation focused engineer} & Simulating large-scale systems & Accurate simulation of physical, chemical systems or dynamical processes & Scalability, approximation quality, solution guarantees & Mid to high: focus on algorithmic accuracy \\\addlinespace\cline{2-6}\addlinespace
            & \emph{Researchers} & Pilot studies, scientific knowledge gain & Scientific discovery, publishing findings & Time constraints, manual transfer of data, resource limitations & Varying: high abstraction for domain specific goal; potentially low for methods\\\addlinespace\cline{2-6}\addlinespace
            & $ \hookrightarrow$ \emph{Physicist} & Verifying theories, exploring condensed matter, simulation & Solving hard physics problems & physical principles, resource limitations & Mid to low: programming languages over quantum instructions, Hamiltonians for physical processes\\\addlinespace\cline{2-6}\addlinespace
            & $\hookrightarrow$ \emph{Chemist} & Solving hard chemistry problems & Scientific / material discovery & chemical principles, resource limitations & Mid: tools / frameworks for chemical simulations, Hamiltonians for chemical processes\\\addlinespace\hline\addlinespace\addlinespace
            \multirow{4}{*}{
            \rotatebox[origin=c]{90}{\makebox[-60pt][r]{\textit{Hardware oriented}}}
            } 
            & \emph{Platform builder} & Building quantum platform (\acrshort{sdk}s, \acrshort{api}s) & Benchmarking quantum hardware / software & Modularity and replaceability of components, lack of standardization in software & Mid: toolchain development and customization \\\addlinespace\cline{2-6}\addlinespace
            & \emph{Quantum\newline algorithm designer} & Designing / improving quantum algorithms & Algorithmic performance (asymptotic \& practical) & Quantum-specific model of computation, lack of standardization & Mid to low: on algorithm interface \& below (needs access to low-level details)\\\addlinespace\cline{2-6}\addlinespace
            & \emph{\acrshort{hpc}\newline engineer} & Enabling heterogenous, large scale computation & System integration, interleaving computation & Ability to interleave communication / coordination & Low: hardware and system integration \\\addlinespace\cline{2-6}\addlinespace
            & \emph{Embedded quantum developer} & Developing scalable and distributed quantum computing  & System-level integration of quantum computing & Interleaving communication/coordination between \Glspl{qpu}, system-specific expertise & Low: focus on hardware / software integration \\
            \bottomrule
    \end{tabular}\par}
\end{table*}

\rqbox{RQ1: Who are the relevant stakeholders in quantum software\, both now and in the future?}{
We propose eleven initial personas for quantum software depicted in bold in \autoref{fig:DagMapsInterview}, which we aggregate based on similarity\footnotemark{}: business users, early adopters, government contractors, researchers, such as physicists and chemists, platform builders, quantum algorithm designers, \Gls{hpc} engineers, embedded quantum developers and simulation focused engineers.
}
\footnotetext{Researcher encompasses a broader scope than physicist or chemist.}

\begin{figure}[tb]
\centering
\includegraphics[width=0.95\linewidth]{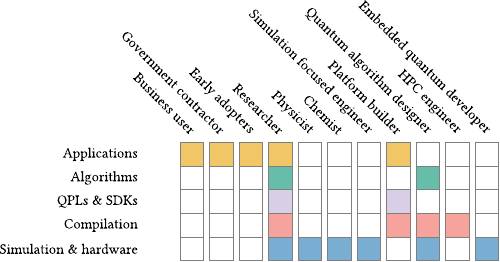}

\caption{The layers of quantum software (y-axis; additionally coloured) in which each identified persona is mostly active.
}
\Description{See text.}
\label{fig:persona_layer}
\end{figure}

To address \textbf{RQ2}, for each aggregated persona, we describe a representative use case, taking into account their interests and constraints as detailed in \autoref{tab:user_groups_zoom}. This data is sourced from expert interviews at Dagstuhl, QCE interviews, and our interpretation.
\autoref{tab:user_groups_zoom} depicts these detailed persona descriptions that are based on the expert interview at Dagstuhl, the interviews at \Gls{qce} and our own interpretation.
These archetypical personas
subsume important characteristics of specialized stakeholders, as they are generalisations from concrete people with concrete use cases and characteristics. 
We group these personas by \emph{application}, and \emph{hardware} orientation
with diverse sets of use cases and applications ranging from industry to specialized technical applications.
Other personas are working on the technical realization of quantum software ---  
requiring hardware awareness.
Similar to the diversity in use cases, interests range from commercial goals and technical objectives to scientific pursuits.
Accordingly, personas are subject to a wide range of constraints, that are, limitations or challenges that stakeholders face when engaging with quantum software.
The mental models of stakeholders influence the abstractions with which they want to interact when using quantum software. 
These range from high-level for application oriented personas to low-level for hardware oriented personas.
In the following, we provide detailed synthesis of the answers to each sub-research question (see \autoref{sec:ResearchAndMethodology}).
\rqbox{RQ2: What are the specific needs of stakeholders in quantum software, and how can abstraction help to reach their goals?}{
We define use cases, interests, constraints and abstraction level for each aggregated persona in \autoref{tab:user_groups_zoom}.
Furthermore, we provide additional insights into abstractions used by personas in \autoref{fig:persona_layer}.
}
\paragraph{RQ2.1}
The application-oriented personas (\eg, business user, early adopter, government contractor and some, high-level, researchers) all share the desire to use quantum computing to solve domain-specific problems more effectively than current classical solutions.
Use cases of personas that are application- and hardware oriented (\ie, different types of researchers and simulation focused engineer) tend to be more specific and technically involved.
For instance, the (physical) simulation of (large-scale) systems requires domain-, as well as simulation-specific details.
Mostly hardware-oriented personas require an even deeper understanding of details that are otherwise abstracted to application-oriented personas:
Quantum algorithm designers improve algorithms at a theoretical and on a practical level, considering the capabilities of hardware.
Platform builders, \Gls{hpc} engineers and embedded quantum developers make a significant contribution to the technical realisation of quantum software and its integration with classical computing.
\paragraph{RQ2.2}
Interests of personas fall into three categories: commercial goals (such as maximizing profit and security), technical objectives (such as benchmarking hardware and optimizing runtime), and scientific pursuits (such as making new discoveries and verifying theories).
Quantum computing is not only a tool that personas apply, but in itself a field of study for all of the aforementioned categories. 
For instance, personas involved in building quantum software are interested in its development and improvement.
\paragraph{RQ2.3}
Personas face various constraints, meaning limitations or challenges, when engaging with quantum technology.
These cover technical limitations imposed by existing machines, as well as the integration of existing interfaces and modularity of software components.
Manual data transfer, academic deadlines and guarantees on software behaviour pose operational challenges.
Constraints such as scalability, adaptability, modularity and replaceability impact the design of (quantum) programs to suit user needs; there may be additional use case-dependent constraints.
For instance, a simulation use case is fundamentally limited (in precision) by scalability, which depends on~--~among others~--~problem structure that might allow for mathematical simplifications or accurate approximations. 
In physics, some Hamiltonians that describe a physical system that allows for scalable simulations.
However, there are a multitude of physical systems requiring different assumptions or modelling techniques to arrive at manageable Hamiltonians, although these use cases are subsumed under the physicist personas.
\paragraph{RQ2.4}
Aforementioned constraints and mental model govern which abstractions personas are interested in.
For instance, application oriented personas do not consider hardware-specific details, yet, these are relevant to determine use-case feasibility.
Similarly, \Gls{hpc} engineers or embedded developers need access to lower-level details to optimise communication and integration;
quantum algorithm designers need to consider (future) capabilities of hardware.

To further illustrate the abstractions required by each persona, we categorize them into five (non-exhaustive) abstraction layers in \autoref{fig:persona_layer}, considering
layers similar to those introduced by \citeauthor{bandicFullstackQuantumComputing2022}~\cite{bandicFullstackQuantumComputing2022}.
Concretely, from high to low levels of abstraction:
end-user \textit{applications};
development of quantum \textit{algorithms};
hardware-agnostic program abstractions such as \textit{\Glspl{qpl}} and \textit{\Glspl{sdk}};
\textit{compilation} of programs into hardware-specific formulations;
\textit{simulation} of quantum-mechanical systems and the development
of quantum \textit{hardware}.
\autoref{fig:persona_layer} shows that certain personas (\ie{}, business users, physicists, chemists, simulation engineers, \Gls{hpc} engineers, and government contractors) mostly work on a single high- or low-level quantum software layer, respectively.
Interestingly, there are personas for which multiple layers are relevant, such as the researcher, the quantum algorithm engineer, and the platform builder.
Therefore, their requirements also span multiple layers.
Since all layers of quantum software\footnote{Specialized personas (see chemist or physicist) may require fewer layers} are of potential interest to the researcher, each must support the ability for rapid (prototypical) development in the short- and mid-term.

\section{Threats to Validity}
\label{sec:ThreatsToValidity}
Despite careful study design, several threats to validity should be considered when interpreting our results.
\paragraph{Participant Selection}
For our focus group, 
we relied on experts who were present at the Dagstuhl seminar (from the authors of this paper, \censor{LaS, JG, RW, SF, IS, WM} contributed to the discussion).
Invitations were limited to pre-selected experts recognized in the field.
However, the attendees of the working group on abstractions were chosen by convenience sampling, since experts had to opt in to participate.
As the objective of this workshop was to develop an initial hypothesis, it seemed reasonable to base the work on the expertise of specialists in the field.
To incorporate the perspective of practitioners 
we conducted interviews
according to the Goal-Directed Design process (see \autoref{sec:Results}).
Interviewer-administered surveys are more social, and therefore at risk of bias due to interviewer variance~\cite{Dillman2024}.
As our interviews were conducted by three different interviewers and three different note-takers, we cannot rule out that interviewer variance and bias from the authors' personal opinions influenced the results, yet deem the risk to be negligible, as all interviewers were aware of this problem, and none has commercial or other interest in specific outcomes.
Due to the limited time available at the conference and to avoid participants' concerns about confidentiality, we opted for manual note-taking rather than audio recording~\cite{flick2023qualitativeResearch}.
To gather a rather diverse set of interviewees, we used convenience sampling at a relevant venue with interest from academia and industry (\Gls{qce}).
Although we deliberately chose \Gls{qce} for its mix of academic and industry visitors, as well as its diversity of nationalities, our choice of venue may introduce bias.
Nevertheless, given the difficulty of sampling from the quantum software community, we believe that our sampling method is reasonable.
We were able to obtain nine interviews using this method, limited by the number of workshop attendees.
There is some indication of data saturation, as multiple personas were mentioned multiple times: For instance, the last interviewed participant did not provide any new personas, but only added to previously established  ones.
\paragraph{Credibility, Resonance, and Transferability }
While we carefully considered established empirical standards~\cite{ralph2020empirical},
quantitative criteria (internal, external and construct validity) do not apply to our qualitative survey.
We evaluate based on credibility, resonance, transferability.
We used a dual approach combining expert knowledge and interviews.
The diverse \Gls{qce} community~\cite{QCE_2026_Attend}
is reflected in demographics and professional focus (see \autoref{fig:demographics_success}), indicating multi-vocality.
As Dagstuhl participants are experts in quantum software engineering and most \Gls{qce} interviewees claim to have significant experience (see~\autoref{fig:demographics_success}), 
we consider their statements credible.
We found agreement between our focus group and  interview findings (see \autoref{sec:discussion}), 
indicating that defined personas resonate with practitioners.
Our personas are focused on quantum software, but coarse-grained enough to allow for a broad range of applications within that field.
Due to the extent of tailoring, we anticipate limited transferability beyond this quantum software scope.

\section{Related Work}

Previous works address challenges~\cite{zhaoQuantumSoftwareEngineering2021, murilloQuantumSoftwareEngineering2025,Yue2023} and application scenarios~\cite{carbonelliChallengesQuantumSoftware2024a} of quantum software.
Personas support practitioners and researchers across domains in the design process~\cite{purittPersonasPracticeAndTheory2003, changPersonasTheoryPractices2008, salminenUseCasesPersonas2022}. 
However, no work has, to the best of our knowledge, defined personas or other user models for quantum software.
We are convinced that the use of personas will benefit the field, help identify and address users' needs, and enable stakeholder integration~\cite{Ammermann2024}.
The use of qualitative methods to address human aspects is increasingly recognised in software engineering~\cite{lenbergQualitativeSoftwareEngineering2024}.
Qualitative methods allow participants to respond more freely, making them suitable for use in exploratory studies~\cite{lenbergQualitativeSoftwareEngineering2024}.
While other studies gather insights from active researchers~\cite{murilloQuantumSoftwareEngineering2025} or conduct semi-structured interviews with quantum software practitioners~\cite{destefanoQuantumSoftwareEngineering2024}, to the best of our knowledge no prior work has conducted systematic, community-level interviews, such as at a quantum software conference.

\section{Conclusion and Outlook}
\label{sec:Conclusion}

In this work, we introduce a persona hypothesis
for quantum software based on the results from an expert focus group at the Dagstuhl seminar 24512 and practitioner interviews at the IEEE Quantum Week (\Gls{qce}) 2025.
We define and discuss 11 aggregated personas of quantum software users and stakeholders regarding their use cases, interests, constraints and abstraction level.
We view our work as the starting point
to consolidate quantum stakeholders at a community level and therefore encourage researchers and practitioners to review and refine user personas introduced in this work.
We encourage the quantum software community to gather further insights from stakeholders through user studies and interviews.
Future surveys can provide additional relevant insights into the specific needs of stakeholders and challenges they face.
We see personas as a useful tool for the design of tailored quantum software products.



\begin{acks}
We thank the organisers of Dagstuhl Seminar 24512, the IEEE \Gls{qce} and
all interviewees for their insightful input.
\blackout{This work was supported by the German Research Foundation, grant 563122858 (MA 9739/1-1 and SCHA 1635/20-1).}
\end{acks}


\bibliographystyle{ACM-Reference-Format}
\bibliography{references}



\end{document}